\documentclass[twocolumn,prb,showpacs,preprintnumbers,amsmath,amssymb]{revtex4}

\usepackage{graphicx}
\usepackage{dcolumn}
\usepackage{bm}
\usepackage{amssymb}
\usepackage{wasysym}
\usepackage[rflt]{floatflt}
\usepackage{bibtopic}

\begin{document}

\preprint{APS}

\title {Quantitative determination of spin-dependent quasiparticle lifetimes and electronic correlations in hcp Cobalt}

   \author{J. S\'anchez-Barriga$^{1,*}$, J. Min\'ar$^2$, J. Braun$^2$, A. Varykhalov$^1$, V. Boni $^3$, I. Di Marco$^{4,5}$, O. Rader$^1$, V. Bellini$^3$, F. Manghi$^3$, H. Ebert$^2$, M. I. Katsnelson$^{5}$,  A. I. Lichtenstein$^6$, O. Eriksson$^4$, W. Eberhardt$^1$, H. A. D\"{u}rr$^1$ and J. Fink$^{1,7}$ \\}

\affiliation {$^1$Helmholtz-Zentrum Berlin f\"{u}r Materialien und Energie,
Elektronenspeicherring BESSY II, Albert-Einstein-Strasse 15, D-12489 Berlin, Germany\\
$^2$ Dep. Chemie und Biochemie, Physikalische Chemie, Universit\"{a}t M\"{u}nchen, Butenandtstr. 5-13, D-81377,M\"{u}nchen, Germany\\
$^3$Dipartimento di Fisica, Universit\`{a} di Modena, Via Campi 213/a, I-41100 Modena, Italy\\
$^4$Department of Physics, Uppsala University, Box 530, SE-751 21 Uppsala, Sweden\\
$^5$ Institute of Molecules and Materials, Radboud University of Nijmejen, Heijendaalseweg 135, 6525 AJ Nijmegen, The Netherlands\\
$^6$Institute of Theoretical Physics, University of Hamburg, 20355 Hamburg, Germany\\
$^7$ Leibniz-Institute for Solid State and Materials Research Dresden, P.O. Box 270116, D-01171 Dresden, Germany}
\date{\today}

\begin{abstract}

We report on a quantitative investigation of the spin-dependent quasiparticle lifetimes and electron correlation effects
in ferromagnetic hcp Co(0001) by means of spin and angle-resolved photoemission spectroscopy. The experimental spectra
are compared in detail to state-of-the-art many-body calculations within the dynamical mean field theory and the three-body
scattering approximation, including a full calculation of the one-step photoemission process. From this comparison we conclude
that although strong local many-body Coulomb interactions are of major importance for the qualitative description of
correlation effects in Co, more sophisticated many-body calculations are needed in order to improve the quantitative
agreement between theory and experiment, in particular concerning the linewidths. The quality of the overall agreement
obtained for Co indicates that the effect of non-local correlations becomes weaker with increasing atomic number.

\end{abstract}
\maketitle

\section{Introduction}

Over the last decades, the electronic structure and the dynamics of electronic states in solids have attracted a lot of attention. A better understanding of the binding energies and quasiparticle lifetimes in correlated systems like 3d ferromagnets \cite{Herring}, transition metal oxides \cite{Wadati-LNP-2007}, 4f rare-earths \cite{Sekiyama-LNP-2007} or high-T$_{c}$ superconductors \cite{Fink-LNP-2007} has been achieved, together with important experimental progress in photoelectron and related spectroscopies.\cite{Hufner-highres} On the theoretical side, the application of density functional theory (DFT) in the local density approximation\cite{Moruzzi-1975} (LDA) has contributed with numerous calculations of single-particle E({\bf k}) band dispersions of solids, surfaces, and ultrathin films. 

In the case of magnetic transition metals, an extensive work in this respect has been carried out over the past thirty years.\cite{Plummer-ACPPRL-1982, Himpsel-AP-1983,  Feder-Polel-1985,Davis-JAP-1986, Kevan-1992, Donath-SSP-1994, SH, Rader-Landolt-1999, Johnson-SPS-2007} However, the question for the exact role of correlation effects in the bandstructure of these systems is still open and no general consensus has been reached due to several causes. The most important reasons for the disagreement between theory and experiments in these metals have been (i) the narrowing of the 3d bands due to correlation effects and (ii) the existence of new features in the photoemission data such as photoemission satellites in the core-level and valence band spectra which are not well described by single-particle approaches such as the generalized gradient approximation\cite{Perdew-PRB-1986} (GGA) or the local spin-density approximation (LSDA) in DFT. 

Although one would expect a better agreement between theory and experiment if the atomic number increases within the 3d series, the case of Co, which is the focus of the present work, is still an intriguing question. First attempts to compare experiment and theory indicated that for Co -but also for Fe- many-body effects can be considered small \cite{Davis-JAP-1986}, in contrast to Ni.\cite{Eberhardt-PRB-1980, Himpsel-PRB-1979, BME+06, MEN+05} While Ni was considered an example of a strongly correlated system where the narrowing of the 3d bands is substantial, first results in Co indicated larger deviations than for Fe\cite{Davis-JAP-1986, Steiner-PRB-92, Alkemper-PRB-1994} where the energy positions with respect to the Fermi energy were about 10\% smaller in the experiment than in LDA.\cite{Santoni-PRB-1991} Moreover, the Fe photoemission peaks showed a large linear broadening ($\sim$60\% of the binding energy),\cite{Santoni-PRB-1991} later on predicted to be accompanied by a strong loss in spectral weight \cite{Katsnelson-JPCM-1999} and most recently attributed to the strong non-locality of correlation effects in Fe.\cite{Sanchez-Barriga-PRL-2009} On the other hand, Co has been the subject of several photoemission studies since the beginning of the 80s \cite{Davis-JAP-1986, Steiner-PRB-92, Himpsel-PRB-1980, Eastman-PRL-1980, Osterwalder-JESRP-2001, Getzlaff-JMMM-1996,Math-SS-2001,Himpsel-CoPRB-1979}, but only a few have been devoted to a detailed comparison between many-body calculations and experiments.\cite{Osterwalder-JESRP-2001, Steiner-PRB-92, Monastra-PRL-2002} Recently, by comparing experimental photoemission spectra of fcc Co to theoretical calculations within the three-body scattering approximation \cite{Calandra-PRB-1994} (3BS), it has been found that due to many-body effects no sharp quasiparticle peaks exist for binding energies larger than 2 eV in this system.\cite{Monastra-PRL-2002} Interestingly, this effect has been theoretically explained by the existence of strong correlation effects which could particularly affect majority spin electrons \cite{Monastra-PRL-2002}, leading to a more pronounced renormalization of the majority spin quasiparticle spectral weight as compared to its minority counterpart. However, these conclusions were not supported by experiments involving spin resolution. On the contrary, in a very recent photoemission experiment on a hcp Co bulk single crystal with out-of-plane magnetization \cite{Mulazzi-PRB-2009} it has been shown that for sufficiently high photon energies (488-654 eV), the spectral peak widths become narrower than previously observed in low-energy experiments and, as a result, quasiparticle bands of weak intensity can be observed even at binding energies larger than 3 eV. This effect has been attributed to the significant reduction of the final-state broadening contribution to the measured linewidths at this photon energy range. In addition, a detailed analysis of the linewidths of the photoemission peaks revealed excellent agreement between experiment and calculations within the dynamical mean-field theory\cite{Grechnev-PRB-2007} (DMFT). Although this conclusion was supported for both majority and minority spin electrons, once again, no spin resolution was provided in these experiments to decompose overlapping contributions of opposite spin. 

In contrast to earlier studies, we utilize in the present paper spin resolution
in order to give an experimental and theoretical corroboration to previous findings. The system under study is an ultrathin hexagonal close packed (hcp) Co film of about 20 monoatomic layers (ML) thickness with an in-plane magnetization.\cite{Fritzsche-JMMM-1995,Getzlaff-JMMM-1996} We compare the results from spin- and angle-resolved photoemission (SARPES) experiments to theoretical DMFT and 3BS many-body calculations, including a complete calculation within the one-step model (1SM) of photoemission. Our main conclusion is that although strong local many-body Coulomb interactions play a major role for the qualitative description of correlation effects in Co, more sophisticated many-body calculations including non-local interactions are needed in order to improve the quantitative agreement between theory and experiment, in particular concerning the linewidths. Although the results are similar to the ones previously reported for Fe,\cite{Sanchez-Barriga-PRL-2009} the better agreement obtained in the case of Co indicates a reduction of the non-locality of correlation effects with increasing atomic number. 

\section{Experiment}

Experiments have been performed at room temperature with a hemispherical SPECS Phoibos 150 electron energy analyzer and linearly polarized undulator radiation at the UE112-PGM1 beamline at BESSY II. For spin analysis, a Rice University Mott-type spin polarimeter has been operated at 26 kV.\cite{Burnett-RSI-94} The angular resolution of the equipment is better than 1$^{\circ}$ and the average energy resolution about 100 meV. The Co(0001) surface has been prepared on W(110) by deposition of 20 ML Co from a high purity wire and post-annealing during 5 min. at $\sim 400\,^{\circ}{\rm C}$. The tungsten substrate was cleaned before deposition by several cycles of annealing in oxygen ($1\times10^{-7}$ mbar) at 1200$^{\circ}{\rm C}$ followed by flashing at 2200$^{\circ}{\rm C}$. The base pressure during the experiments was 1--$2\times10^{-10}$ mbar. Figure 1(a) shows an overview of the geometry of the experiment. The linear polarization of the incident photon beam has been horizontal or vertical, and its angle of incidence with respect to the surface normal $\sim$45$^{\circ}$. Hence, in the reference frame of the sample, the light had more p- or full s-character, respectively. The structural quality of the Co films has been checked by low-energy electron diffraction (LEED).
\begin{figure}[t!]
\vspace{0cm}
\centering
\includegraphics [width=0.4\textwidth]{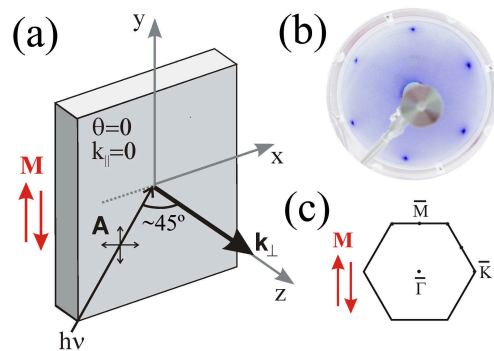}
\caption{(color online). (a) Geometry of the experiment. (b) Six-fold symmetry observed in the LEED pattern of the hcp Co(0001) surface. (c) Sketch of the SBZ of the hcp lattice. In (a) and (c), the magnetization orientation is also indicated.}
\label{fig:Fig1}
\end{figure}
Fig. 1(b) shows the measured LEED pattern of the hcp Co(0001) surface acquired with an incident electron beam of $\sim$150 eV. Note that each diffraction spot corresponds to a $\overline{\Gamma}$ point of the surface Brillouin zone (SBZ), sketched in Fig. 1(c). The LEED pattern of Fig. 1(b) confirms that the annealing procedure which follows the epitaxial growth leads to a well-ordered surface of very high quality. It is well-known that in hcp Co(0001), the magnetization orientation lies out-of-plane in bulk single crystals \cite{Mulazzi-PRB-2009} and in-plane due to surface anisotropy effects in ultrathin epitaxial films grown on W(110).\cite{Fritzsche-JMMM-1995,Getzlaff-JMMM-1996} Therefore, in our experiments the sample was remanently magnetized in the film plane along the Co[1$\overline{1}$00] (or W[1$\overline{1}$0]) easy axis \cite{Rampe, Math-SS-2001, Bettac}, as indicated in Fig. 1(a) and Fig. 1(c). Note that this direction is parallel to the $\overline{\Gamma}$-$\overline{M}$ symmetry line of the SBZ.

\section{Theory}
\label{sec:theory}

We start our considerations by a discussion of Pendry's formula for the photocurrent which defines the one-step
model of photoemission \cite{Pen76}:
\begin{equation}
  I^{\rm PES} \propto
  {\rm Im}~
  \langle \epsilon_f, {\bf k}_{\|} |
  G_{2}^+ \Delta G_{1}^+ \Delta^\dagger G^-_{2} |
  \epsilon_f, {\bf k}_{||} \rangle \: .
\label{eq:pendry}
\end{equation}
The expression can be derived from Fermi's golden rule for the transition probability per unit time.\cite{Bor85} Consequently, $I^{\rm PES}$ denotes the elastic part of the photocurrent. Vertex renormalizations are neglected.
This excludes inelastic energy losses and corresponding quantum-mechanical interference terms.\cite{Pen76,Bor85,CLRRSJ73} Furthermore, the interaction of the outgoing photoelectron with the rest system
is not taken into account. This ``sudden approximation'' is expected to be justified for not too small
photon energies. We consider an energy-, angle- and spin-resolved photoemission experiment.
The state of the photoelectron at the detector is written as $|\epsilon_f, {\bf k}_{\|} \rangle$,
where ${\bf k}_{\|}$ is the component of the wave vector parallel to the surface, and $\epsilon_f$ is the
kinetic energy of the photoelectron. The spin state of the photoelectron is implicitly included in
$|\epsilon_f, {\bf k}_{\|} \rangle$ which is understood as a four-component Dirac spinor. The advanced
Green function $G_{2}^-$ in Eq.~(\ref{eq:pendry}) characterizes the scattering properties of the material at the final-state
energy $E_2 \equiv \epsilon_f$. Via $|\Psi_f \rangle = G^-_{2}|\epsilon_f, {\bf k}_{\|}\rangle$ all
multiple-scattering corrections are formally included. For an appropriate description of the photoemission
process we must ensure the correct asymptotic behaviour of $\Psi_f({\bf r})$ beyond the crystal surface,
i.e. a single outgoing plane wave characterized by $\epsilon_f$ and ${\bf k}_\|$. Furthermore, the damping
of the final state due to the imaginary part of the inner potential $V_{0{\rm i}}(E_2)$ must be taken into
account. We thus construct the final state within spin-polarized low-energy electron diffraction (SPLEED)
theory considering a single plane wave $|\epsilon_f,{\bf k}_\|\rangle$ advancing onto the crystal surface.
Using the standard layer-KKR method \cite{KKR} generalized for the relativistic case \cite{Bra96,Bra01},
we first obtain the SPLEED state $U \Psi_f({\bf r})$. The final state is then given as the time-reversed
SPLEED state ($U=-i \sigma_y K$ is the relativistic time inversion). Many-body effects are included
phenomenologically in the SPLEED calculation, by using a parameterized, weakly energy-dependent and complex
inner potential $V_0(E_2)=V_{0{\rm r}} (E_2)+iV_{0{\rm i}}(E_2)$. The imaginary part $V_{0{\rm i}}(E_2)$ was chosen constant in energy. The best agreement between theory and experiment was obtained for a value of 2.0 eV for all excitation energies, which means it was sufficient to take into account 20 layers for the photocurrent calculation. For the quantum number l we used a maximum value l$_{max}$=3 to account for the contribution of d-f like transitions in the photoemission matrix elements.\cite{Pen74} This generalized inner
potential takes into account inelastic corrections to the elastic photocurrent \cite{Bor85} as well as the actual
(real) inner potential, which serves as a reference energy inside the solid with respect to the vacuum level.\cite{HPM+95} Due to the finite imaginary part $V_{0{\rm i}}(E_2)$, the flux of elastically scattered
electrons is permanently reduced, and thus the amplitude of the high-energy wave field $\Psi_f({\bf r})$ can
be neglected beyond a finite distance from the surface. In order to take into account strong electronic
correlation effects in the initial states we use the LSDA+DMFT approach realized in the framework of the fully
relativistic Korringa-Kohn-Rostoker multiple scattering theory\cite{Min05} (SPRKKR). The corresponding self-energy
${\Sigma}^{DMFT}(E)$ is calculated fully self-consistently (e.g.\ in charge and self energy) via the DMFT approach.\cite{GKKR96} Some of the results have been cross checked by non-self consistent LSDA+DMFT calculations, implemented
within the full potential linear muffin tin orbital (FP-LMTO) method.\cite{Grechnev-PRB-2007} This allows in particular to
use the full-potential option which might be important for more open structures. As a DMFT-solver the relativistic
version of the so-called Spin-Polarized T-Matrix Plus Fluctuation Exchange (SPTF) approximation \cite{KL02,PKL05} was
used. In the present study, all the calculations were performed with a bulk-like self-energy, i.e. not layer-dependent. However, corresponding Bloch spectral functions were cross-checked with the FP-LMTO results using layer-dependent self-energy and only weak quantitative differences were found.  

In addition to these approaches we accounted for correlation effects within the three-body scattering approximation.\cite{Calandra-PRB-1994} Within the 3BS-approach the self-energy is calculated using a configuration interaction-like expansion. In particular three-particle configurations like one hole plus one electron-hole pair are explicitly
taken into account within 3BS-based calculations. The corresponding output can be directly related to the
photoemission process and allows for a detailed analysis of various contributions to the self-energy
(e.g., electron-hole lifetime).     

\section{Results and discussion}

In this section, we discuss the effects of electron correlations in the spin-dependent electronic structure of hcp Co. In subsection A, the relativistic notation used to label the symmetries of the observed states will be introduced. In subsection B, a detailed analysis of the different transitions observed in the experimental and theoretical SARPES spectra will be given, focusing on the quality of the agreement between theory and experiment. Finally, in subsections C and D, further discussion on the many-body aspects of the Co bandstructure will be presented, together with a quantitative analysis of the experimental and theoretical linewidths of the Co photoemission peaks.  

\subsection{Selection rules and hcp notation} 
\begin{figure*}[tp]
\centering
\includegraphics [width=0.65\textwidth]{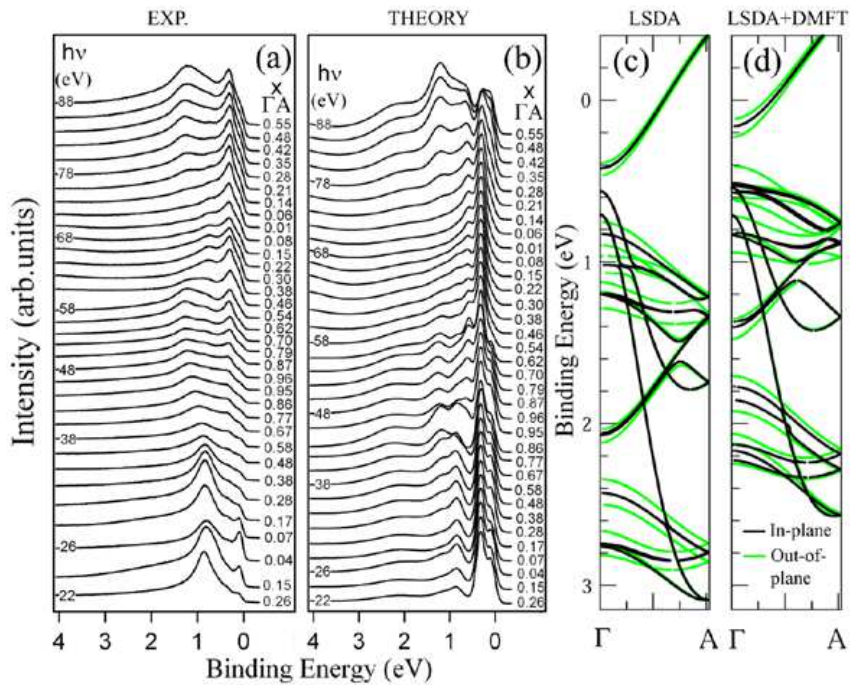}
\caption{(color online). (a) Experimental spin-integrated photoemission spectra of hcp Co(0001) measured with p-polarization
in normal-emission along the $\Gamma$A direction of the bulk Brillouin zone at different excitation energies (steps of 2 eV). The curves are labeled by selected photon energies and by the wave vectors in units of
$\Gamma$A=0.77$\AA^{-1}$ ($\Gamma$ point at x=0 and A point at x=1). (b) Corresponding calculation obtained by the LSDA+DMFT+1SM
method for hcp Co with an in-plane magnetization along the $[1\overline{1}00]$ direction. (c) and (d): Spin-integrated band
structure calculations for two configurations of the magnetization, out-of-plane (along $[0001]$ direction, green (light) lines)
and in-plane (along $[1\overline{1}00]$ direction, black (dark) lines). In (c), LSDA calculations are shown, and in (d) fully
relativistic valence band states obtained self-consistently in the SPRKKR formalism by suppressing the imaginary part of the
self-energy Im$\Sigma_{DMFT}$.}
\label{fig:Fig2}
\end{figure*}
In contrast to previous studies of hcp Co,\cite{Himpsel-PRB-1980} where the non-relativistic notation for fcc crystals was used to label the symmetry of the observed states, from now on we will follow the relativistic notation given by Benbow \cite{Benbow-PRB-1980} for hcp crystals. This notation is much simpler since due to the lower symmetry it only involves initial and final states of $\Delta_{7}$, $\Delta_{8}$ and $\Delta_{9}$ symmetries. This means that in the geometry used in the experiment (Fig. 1(a)) and due to selection rules, only transitions of $\Delta_{7}$ and $\Delta_{8}$ symmetries are suppressed for s-polarization when moving from $\Gamma$ to A and from A to $\Gamma$ in the next Brillouin zone, respectively. Hence, only transitions of $\Delta_{9}$ symmetry can be observed with s-polarized photons. The effect of matrix-elements, which is also influenced by the symmetry of the hcp crystal structure\cite{Aschcroft}, is similar to the interference effect or Brillouin zone-selection rule reported in Ref. \onlinecite{Shirley-PRB-1995}, which suppresses the intensities of certain bands in a particular Brillouin zone but increases them in the next-higher one. The overall effect leads to a bandstructure which is equivalent to the one of an fcc crystal if all the hcp bands are folded back about the A point.\cite{Himpsel-PRB-1980} This also implies that states of $\Delta_{7}$ and $\Delta_{8}$ symmetry should be interchanged when crossing the A point while states of $\Delta_{9}$ symmetry remain unaffected. To avoid redundancy, from now on double subscripts will be adopted, unless specified. For better clarity, the subscript notation of almost degenerated states will follow their order of appearance from lower to higher binding energy (BE).

\subsection{Spin and angle-resolved photoemission: a comparison theory-experiment}

Figures 2(a) and 2(b) show a comparison between the spin-integrated experimental spectra of hcp Co(0001) and the corresponding theoretical LSDA+DMFT+1SM calculations for p-polarization along the $\Gamma$A direction of the bulk Brillouin zone. The {\it k} values were calculated from the measured photon energies, ranging from 22 to 80 eV and using an inner potential $V_{0}$=14.8 eV. The best correspondence between the BE positions of the experimental and theoretical peaks was found for values of the averaged on-site Coulomb and exchange interactions of U=2.5 eV and J=0.9 eV, in agreement with previous theoretical and experimental studies.\cite{Grechnev-PRB-2007,Monastra-PRL-2002} Therefore, unless specified, these have been the parameters of choice in all the calculations presented here. 

Figures 2(c) and 2(d) show LSDA and fully-relativistic LSDA+DMFT calculations of the spin-integrated band structures for
different directions of the sample magnetization, the out-of-plane [0001] and in-plane $[1\overline{1}00]$ directions
(green (light) and black (dark) lines, respectively). It can be observed how the different bands shift between Fig. 2(c) and
Fig. 2(d) towards lower binding energies due to electron correlation effects for both magnetization orientations. 

As already mentioned above, the Co(0001) films are magnetized in-plane along the $[1\overline{1}00]$ direction.
All LSDA+DMFT+1SM calculations presented here have been performed using in-plane magnetization direction. However,
we show in addition out-of-plane calculations in order to emphasize how due to spin-orbit coupling effects, changes in the orientation
of the magnetization correspond to changes in the electronic structure of hcp Co. This purely relativistic effect \cite{Ackermann-JPFMP-1984, Bra01, Braun-JPCM-2004} may gain importance if e.g., experimental and theoretical spectra of hcp Co bulk single crystals \cite{Himpsel-PRB-1980, Mulazzi-PRB-2009} and overlayers \cite{Getzlaff-JMMM-1996, Monastra-PRL-2002} need to be compared. To a first approximation, the magnetocrystalline anisotropy energy is given by the average energy difference between the BE positions of the bands in Fig. 2(c) and Fig. 2(d) with in-plane and out-of-plane magnetization directions, respectively. The overall effect of the magnetic anisotropy on the bandstructure results in slight changes in the dispersion behaviour of some bands, additional BE shifts and most importantly, a considerable reduction of the spin-orbit splitting with an in-plane magnetization configuration. In this case, near the Fermi level and below 2 eV, most of the bands are degenerated while in the out-of-plane configuration, the spin-orbit splitting is $\sim$0.1 eV in average, a value of the order of the experimental energy resolution. Therefore, it should be emphasized that relatively narrower peaks would be expected in photoemission experiments with an in-plane configuration of the magnetization, such as in Co overlayers as opposed to hcp Co bulk single crystals. 

\begin{figure}[tbp]
\vspace{0.5cm}
\centering
\includegraphics [width=0.4\textwidth]{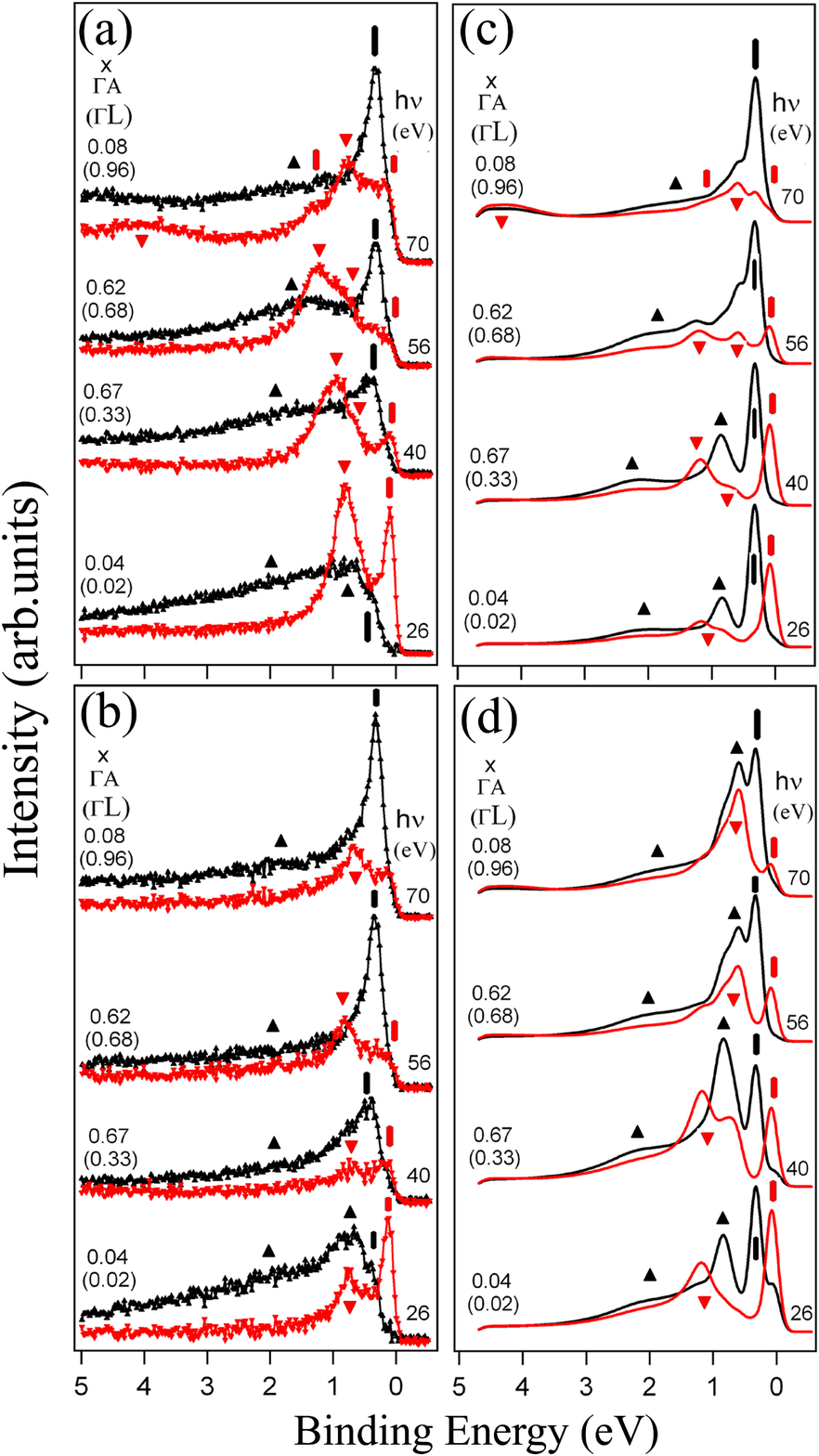}
\caption{(color online). Spin-resolved spectra of hcp Co(0001) for different polarizations. (a), (b) Experiment (upwards (black (dark)) triangles: majority spin states; downwards (red (light)) triangles: minority spin states). (c), (d): LSDA+DMFT+1SM theory (black (dark) and red (light) lines for majority and minority spin electrons, respectively). (a),(c) for p- and (b), (d) for s-polarization. Additional upwards (black) [downwards (red)] triangles on top of the spectra indicate majority [minority] bulk states, while vertical black (red) bars majority (minority) spin surface-related features. The curves are labeled by the photon energies (right) and the wavevectors (left) for both the hcp ($\Gamma$A=0.77$\AA^{-1}$) and fcc ($\Gamma$L=1.54$\AA^{-1}$) lattices for comparison.}
\label{fig:Fig3}
\end{figure}
Let us now focus on the origin of the different transitions appearing in the theoretical and experimental spectra of Fig. 2 and Fig. 3, from lower to higher binding energies. In Fig. 3, similar data to the ones presented in Fig. 2(a) and Fig. 2(b) are shown, but now at a few selected photon energies and spin resolved. Figures 3(a) and 3(b) show experimental SARPES spectra for p- and s-polarized photons, respectively, while Fig. 3(c) and Fig. 3(d) show the corresponding LSDA+DMFT+1SM calculations. 
All the peaks appearing near the Fermi level in the full photon energy range and for a BE lower than $E_{B}\sim$0.5 eV, can be attributed to the two majority and minority spin components of a Tamm-like surface feature, in agreement with previous experimental observations.\cite{Himpsel-CoPRB-1979} Closest to the Fermi level, the minority spin component of this state appears at $E_{B}\sim$0.05 eV, and it can be identified as a pure surface state of $\Delta_{9}^{\downarrow}$ symmetry, since it lies within the gap appearing near the Fermi level and its intensity is nearly the same for both p- and s-polarized photons (see Fig. 3). Only near the first $\Gamma$ point and at x$\sim$0.04$\Gamma$A (h$\nu$=26 eV), this peak is slightly influenced by the minority spin bulk-like band of $\Delta_{9,7}^{\downarrow}$ symmetry crossing the Fermi level. Thus, only at this photon energy it acquires a slight bulk-like character and it can be identified as a surface resonance feature. 

A similar argumentation can be used for the majority spin component of the abovementioned surface-like Tamm state, which can be identified as a surface resonance feature of $\Delta_{9,7}^{\uparrow}$ symmetry (or equivalently, $\Delta_{9,8}^{\uparrow}$ symmetry in the next Brillouin zone), since it is located at the border of the gap and almost degenerated with the majority bulk bands appearing around the same BE. This state, which appears at $E_{B}\sim$0.4 eV, can be observed as a non-dispersing feature in the full photon energy range (see Fig. 2(a), Fig. 2(b) and Fig. 3). However, when moving in the experiment from $\Gamma$ to A in a photon energy range between 26 and 38 eV (0.04$\leq$ x $\leq$ 0.58) (Fig. 2(a)), its intensity is much weaker as compared to the theoretical calculations (Fig. 2(b)). This can also be noticed in the results of Fig. 3 near the $\Gamma$ point (x$\sim$0.08 $\Gamma$A, h$\nu$=26 eV) for both p and s polarizations. In this case it can be observed as a majority spin shoulder close to the Fermi level, but in the theoretical spectra as a sharp peak. Since in the calculations its intensity does not seem to be strongly influenced by matrix-element effects and in addition it exhibits a significantly narrower linewidth, we attribute these strong differences to a theoretical underestimation of the multiple scattering events occurring between the surface and the bulk electron wavefunctions. Moreover, the intensity of this peak is strongly enhanced at higher energies in the experiment (see Fig. 2(a), Fig. 3(a) and Fig. 3(b)), particularly when moving from around the A point (x$\sim$0.98$\Gamma$A, h$\nu$=48 eV) to the second $\Gamma$ point (x$\sim$0.01$\Gamma$A, h$\nu$=72 eV) and beyond. This increase in intensity can be attributed, on the one hand, to a slightly reduced surface sensitivity at higher photon energies, and on the other hand, to the contribution of the nearly-flat majority spin bulk-like band of $\Delta_{9,8}^{\uparrow}$ symmetry appearing around the same BE for photon energies h$\nu>$48 eV. Note once again, that due to the symmetry arguments mentioned above, for h$\nu\leq$48 eV this bulk band has $\Delta_{9,7}^{\uparrow}$ symmetry and is located at a BE of about 0.7 eV near the first $\Gamma$ point (h$\nu$=26 eV) (Fig. 3). 

All the other bands that are left to describe are bulk-like and appear at a BE higher than $\sim$0.5 eV for all the measured {\it k} points. Firstly, we observe a minority spin band of $\Delta_{9,7}^{\downarrow}$ symmetry located in the experiment at $E_{B}\sim$0.8 eV in the first $\Gamma$ point (see Fig. 2(a), Fig. 3(a) and Fig. 3(b)). Through the polarization dependence of the experimental spin-resolved data of Fig. 3(a) and Fig. 3(b), it is possible to see that this band comprises several components which are all degenerated at the first $\Gamma$ point. These components slightly disperse when the photon energy is increased, giving rise to shoulders which can be detected through polarization-dependent analysis. Most prominently, at x=0.67$\Gamma$A (h$\nu$=40 eV) a minority spin shoulder at a BE of about 0.75 eV can be clearly distinguished in the lower BE energy side of the peak appearing at $E_{B}\sim$0.95 eV for p-polarization data (Fig. 3(a)). Detailed analysis reveals that the BE position of this shoulder exactly coincides with the minority spin peak of $\Delta_{9}^{\downarrow}$ symmetry appearing in the corresponding spectrum for s-polarization (Fig. 3(b)). Further information on the symmetry of these states can be obtained from a comparison to the spin-resolved theoretical spectra of Fig. 3(c) and Fig. 3(d). In particular, it can be demonstrated that the peak appearing at 0.95 eV must be assigned to a state of $\Delta_{7}^{\downarrow}$ symmetry, and the shoulder to a state of $\Delta_{7,9}^{\downarrow}$ symmetry. A similar effect occurs at higher photon energies in the next Brillouin zone e.g., at x=0.62$\Gamma$A (h$\nu$=56 eV), where the peak at $E_{B}\sim$1.2 eV and the shoulder at $E_{B}\sim$0.8 eV can be assigned to bands of $\Delta_{8}^{\downarrow}$ and $\Delta_{8,9}^{\downarrow}$ symmetries, respectively. In the next $\Gamma$ point and at x=0.08$\Gamma$A (h$\nu$=70 eV), the peaks show the same behaviour, with the difference that a weak shoulder appears at $E_{B}\sim$1.5 eV, exactly in the higher BE side of the bulk-like $\Delta_{8,9}^{\downarrow}$ state located at $E_{B}\sim$0.8 eV. Only by means of a detailed comparison to the theoretical spectra of Fig. 3(b), we can attribute this feature to a $\Delta_{8,9}^{\downarrow}$ surface resonance. The fact that it was not visible at lower photon energies is possibly related to the dispersion of the bulk-like minority spin peaks we have just discussed. Besides, this may also be related to the very short lifetime of surface resonance features at higher BE caused by the strong increase of Im$\Sigma_{DMFT}$, which also leads to a very large broadening and makes it impossible to properly distinguish them in the experiment.

At last, we are only left with the majority spin bulk-like states at higher BE. These peaks are strongly broadened and more difficult to distinguish in the experiment, particularly in the spectra of Fig. 2(a) due to the scaling of the intensities. However, with the use of spin resolution (see Fig. 3(a) and Fig. 3(b)), it can be seen that they are mostly suppressed with s-polarization at binding energies around 2 eV. These states become clearly visible in a BE range from about 1.4 to 1.9 eV for p-polarization data in Fig. 3(a) at x=0.67$\Gamma$A (h$\nu$=40 eV) and x=0.62$\Gamma$A (h$\nu$=56 eV), and for s-polarization data in Fig. 3(b) near the first $\Gamma$ point at x=0.04$\Gamma$A (h$\nu$=26 eV). In the calculations (Fig. 3(c) and Fig. 3(d)), on the other hand, the BE of these states seems to be well reproduced, but they are more pronounced in intensity. From the polarization dependent analysis of the experimental data and by comparison to the theoretical calculations, the assignment of the symmetries in this case can be derived from selection rules in a straightforward manner. Near the first $\Gamma$ point (h$\nu$=26 eV), where all the states of this kind are degenerated, we can observe them with both polarizations, and thus the dominant symmetry should be $\Delta_{9}^\uparrow$. At x=0.67$\Gamma$A (h$\nu$=40 eV), however, the dominant symmetry should be $\Delta_{7}^\uparrow$, since the peaks are mostly suppressed with s-polarization. Moving into the next Brillouin zone, the same argument applies and therefore, the dominant symmetries are $\Delta_{8}^\uparrow$ and $\Delta_{8,9}^\uparrow$ for x=0.62$\Gamma$A (h$\nu$=56 eV) and 0.08$\Gamma$A (h$\nu$=70 eV), respectively.

Let us now discuss the quality of the agreement between the theoretical and experimental spectra shown in Fig. 2 and Fig. 3. The first observation is that we find rather good agreement for most of the BE positions of the peaks appearing in the spectra. The second observation, concerning the linewidths, is that the larger broadening of the experimental peaks indicates that the theory underestimates the scattering rates, similar to what has been found in Fe.\cite{Sanchez-Barriga-PRL-2009} The third observation, in contrast to recent experimental conclusions \cite{Mulazzi-PRB-2009}, is that no quasiparticle bands derived from 3d states appear at binding energies higher than $\sim$2 eV. At around 2 eV, this effect is even more pronounced in the experiment than in the theory due to the increased linewidth broadening mentioned above. This finding is in agreement with previous experimental and theoretical analysis in the framework of the 3BS theory \cite{Monastra-PRL-2002}, in which the quenching of the majority spin channel quasiparticle excitations was identified as the main mechanism for this intriguing effect. Our results corroborate this conclusion in view of the spin-resolved experimental and theoretical spectra presented in Fig. 3. Only peaks with strong sp character are observed at higher BE energies than 2 eV, such as the minority spin sp-band of $\Delta_{8}^{\downarrow}$ symmetry appearing at x$\sim$0.08 $\Gamma$A (h$\nu$=70eV) for p-polarization in Fig. 3(a) and in Fig. 3(c) at a BE of about 4.3 eV. 

\begin{figure}[b]
\includegraphics [width=0.50\textwidth]{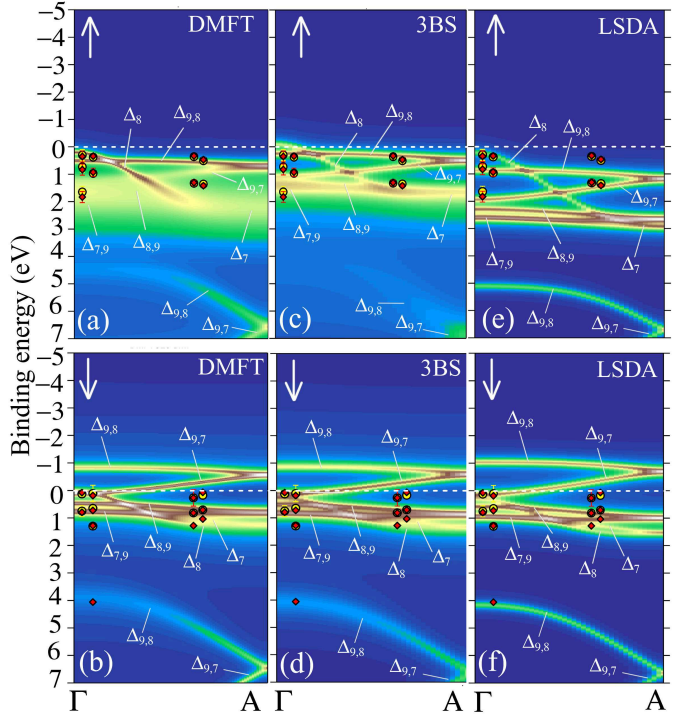}
\caption{(color online). Quasiparticle spectral functions of hcp Co and the photoemission peak positions obtained from the spin-resolved measurements of Fig.~\ref{fig:Fig3} for different polarizations ($\Diamond$ for p- and $\ocircle$ for s- polarization). The results are obtained by the LSDA+DMFT ((a),(b)), the LSDA+3BS ((c),(d)) and the LSDA ((e),(f)) methods for majority ((a),(c),(e)) and minority spin ((b),(d),(f)) electronic states, respectively. The symmetries of the different states are indicated.}
\label{fig:Fig4}
\end{figure}
\subsection{Many-body aspects of the bandstructure}

Comparing the experimental results from spin-integrated and spin-resolved ARPES measurements with LSDA+DMFT+1SM results, good agreement is obtained for most of the peak positions. This is also shown in Fig. 4, where the BE positions of the different experimental peaks we have just described are compared to the calculated bulk-like spin-dependent spectral functions A({\bf k},E) obtained in the framework of the LDSA+DMFT (Fig. 4(a) and Fig. 4(b)), LSDA+3BS (Fig. 4(c) and Fig. 4(d)) and LSDA (Fig. 4(e) and Fig. 4(f)) schemes. Majority spin states are shown in Fig. 4(a), Fig. 4(c) and Fig. 4(e) and minority spin states in Fig. 4(b), Fig. 4(d) and Fig. 4(f). The symmetry labels of the states are also given. In these calculations, the experimental energy resolution is considered by a broadening of $\sim$100 meV. In a first approximation, Figs. 4(a)-4(d) give a comparison between the spin-dependent experimental Re$\Sigma$ and theoretical Re$\Sigma_{DMFT}$ and Re$\Sigma_{3BS}$. However, it should be noted that the spectral functions A({\bf k},E) contain more bands than the ones observed in the experiment, since Brillouin zone selection effects are not considered in this case. This means that in these calculations matrix-element effects, surface effects, and the actual steps of the photoemission process (i.e., propagation of the photoelectron to the crystal surface and its escape into the vacuum) are not included. The experimental peak positions were extracted from the experimental SARPES data of Fig. 3(a) and Fig. 3(b) by means of a fitting procedure which will be described in more detail below. Before proceeding, it should be noted that from Figs. 4(a)-4(d), it is clear that the theoretical results derived from both LSDA+DMFT and LSDA+3BS methods are well in agreement with each other for identical values of U and J. 

Let us focus first on majority spin states. From a first examination across the complete Brillouin zone, one can conclude that the LSDA+DMFT and LSDA+3BS spectral functions reproduce the experimental data better than standard LSDA calculations. This is most apparent at lower BE and for states near the A point, where the LSDA bands are shifted to higher BE as compared to the experimental points assigned to majority spin states of $\Delta_{9,8}^{\uparrow}$ symmetry, as described above. Thus, in order to find best agreement with the experiment, we would need to shift  by about 0.5 eV the corresponding majority spin LSDA bands towards lower BE. Moreover, note the remarkable intensity of the LSDA majority spin bands appearing in a BE range from 2.5 to 3 eV in the whole Brillouin zone. As discussed before, these bands are strongly broadened in the experiment and appear at BE energies of around 1.7-1.9 eV. From the polarization-dependent analysis carried out before, we also know that at the different k-points of the bandstructure, the symmetries of these states  [$\Delta_{9}$($\Delta_{8,9}$) and $\Delta_{7}$($\Delta_{8}$)] coincide with the symmetries of the LSDA bands appearing at $E_{B}>$2 eV, which are not forbidden by selection rules. This means that in Fig. 4(e), the agreement between the peak positions from the experimental peaks around 1.9 eV with the LSDA bands of $\Delta_{9,7}^{\uparrow}$ appearing at binding energies below 2 eV is just accidental, since this agreement does not hold based on simple symmetry arguments. Therefore, also in this case a shift of these states by about 0.7-1 eV becomes necessary to properly explain the bandwidth reduction observed in the experiment. These two facts are already an indication that a narrowing of the 3d bands due to correlation effects becomes necessary to properly describe the experimental data. This is clearly demonstrated in Fig. 4(a) and Fig. 4(c), where a relatively good agreement is obtained for the BE positions of the experimental majority spin peaks for both p and s polarizations. In this case, a reduction of the bandwidth of majority spin states by $\sim$37\% is necessary, in agreement with the results presented in Fig. 2(c) and Fig. 2(d). 

Regarding the minority spin states, it is clear that correlation effects are weaker than for majority spin. Note that close to the A point, the $\Delta_{9}^{\downarrow}$ minority spin component of the surface-like Tamm state discussed above is also visible. This state is not reproduced in the spectral function calculations because, as already explained, surface effects are not included in these calculations. In fact, the different minority spin bands in the LSDA+DMFT (Fig. 4(b)) and LSDA+3BS (Fig. 4(d)) calculations shift by about 0.1 eV respect to the LSDA ones (Fig. 4(e)). This means that the BE positions of the different bands in the three different theoretical approaches show relatively good agreement with the experimental data. This is something expected a priori, since from the theory we know that correlation effects should be more pronounced for majority spin electrons than for minority spin ones, which is a common feature for 3d transition metals. This can be easily explained by considering the creation of electron-hole pairs in the minority spin channel: since there are less minority spin electrons and thus more empty states of this kind, for any processes involving electron-hole pair creation, the pair is more likely to appear in the minority spin bands. This means that any scattering process involving majority spin electrons mostly leads to the creation of minority spin electron-hole pairs, with an effective interaction in the complete process proportional to U. On the other hand, scattering processes involving minority spin electrons also lead to the creation of minority spin electron-hole pairs, but the effective interaction for parallel-spin pairs is proportional to U-J$<$U. Therefore, correlation effects are stronger for majority spin electrons and no big differences are observed between the results presented in Fig. 4(b), Fig. 4(d) and Fig. 4(f). 

\begin{figure}[t]
\includegraphics [width=0.48\textwidth]{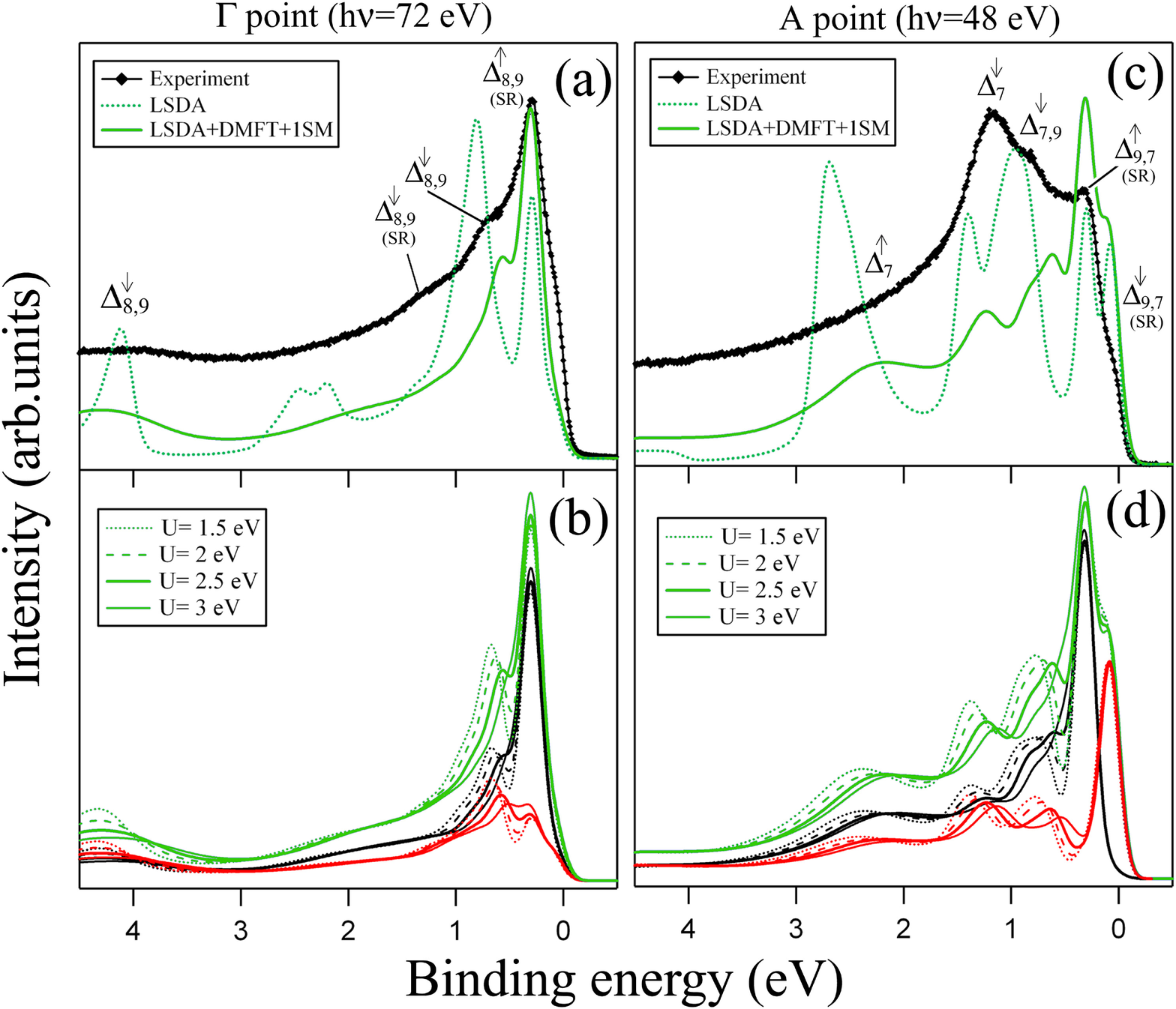}
\caption{(color online). Comparison between experimental and theoretical spectra at the $\Gamma$ [(a) and (b), h$\nu$=72 eV] and A [(c) and (d), h$\nu$=48 eV] points for p-polarization. (a) Spin-integrated experimental spectra [thick (black) symbols], single-particle LSDA-based calculation including surface effects [thick dotted (green) line] and LSDA+DMFT+1SM spectra [thick solid (green) line]. (b) Spin-dependent LSDA+DMFT+1SM calculations for different U values. Black (dark) [red (light)] color denote majority [minority] spin and green (lighter) color the corresponding spin-integrated spectrum. Thin dotted, dashed, thick and thin solid lines for U=1.5, 2, 2.5 and 3 eV, respectively. In (c) and (d), analogous data as in (a) and (b). Symmetry labels in the two consecutive Brillouin zones of the hcp lattice are also indicated.}
\label{fig:Fig5}
\end{figure}

The next point that we should discuss is the agreement between the experimental and theoretical LSDA+DMFT+1SM spectra as a function of the on-site Coulomb interaction U. As already mentioned, the best agreement between theory and experiment was found for U=2.5 eV and J=0.9 eV. The value of J adopted here can be considered a good general assumption for all 3d elements, since J is only a weakly screened parameter. This means that no differences are observed in the LSDA+DMFT+1SM spectra for different J values. Therefore, the only parameter directly linked to the effect of electronic correlations which might lead to changes in the calculated spectra is U and we will proceed with discussing how sensitive the theory is to this parameter in view of the experimental results. 

In Fig. 5, we compare the spin-integrated experimental spectra measured for p-polarization at the $\Gamma$ point (x=0.01$\Gamma$A, h$\nu$=72 eV) [Fig. 5(a)] and near the A point (x=0.96$\Gamma$A, h$\nu$=48 eV) [Fig. 5(c)] to the LSDA+DMFT+1SM and LSDA calculations broadened by the experimental energy resolution. In this case, the LSDA calculations also include surface effects for a better comparison. Among the two spectra available in the data near the $\Gamma$ point, the spectrum at h$\nu$=72 eV has been selected as a suitable choice because of the better agreement at higher photon energies between the experimental and theoretical intensities of the majority spin $\Delta_{8,9}^{\uparrow}$ surface resonance located near the Fermi level. The spin character and symmetry of the different experimental states are also given. In Fig. 5(b) and Fig. 5(d), spin-dependent LSDA+DMFT+1SM calculations for U ranging from 1.5 to 3 eV are also shown. 

In general, the Co LSDA+DMFT+1SM calculations lead to an important improvement as compared to the LSDA calculations. In both Fig. 5(a) and Fig. 5(c), the BE positions of the theoretical peaks agree well with the experimental ones if correlation effects are included. Regarding majority spin states, in Fig. 5(a) a good agreement is achieved for the BE position of the $\Delta_{8,9}^{\uparrow}$ surface resonance near the Fermi level for both the LSDA+DMFT+1SM and LSDA calculations. The bulk component of this peak, which appears at $E_{B}\sim$0.78 eV in the LSDA calculations, is shifted due to correlation effects in the LSDA+DMFT+1SM calculation to $E_{B}\sim$0.47 eV (see Fig. 5(b), U=2.5 eV). In the experiment these peaks are almost degenerated at that BE due to the experimental energy resolution. A similar situation can be found in Fig. 5(c), where good agreement is obtained in the BE position of the $\Delta_{9,7}^{\uparrow}$ majority spin component of the Tamm-like surface state.  

Let us focus now on the most prominent bulk-like minority spin peaks appearing in the theoretical and experimental spectra. The $\Delta_{8,9}^{\downarrow}$ minority spin bulk-like state, which in the LSDA calculation is located at a $E_{B}\sim$0.8 eV, is shifted to $E_{B}\sim$0.66 eV in the experiment (Fig. 5(a)) and to $E_{B}\sim$0.57 eV in the LSDA+DMFT+1SM calculation (Fig. 5(b)) due to correlation effects. The $\Delta_{7,9}^{\downarrow}$ minority spin bulk-like state, on the other hand, is shifted to $E_{B}\sim$0.79 eV in the experiment (Fig. 5(c)) and to $E_{B}\sim$0.63 eV in the LSDA+DMFT+1SM calculation (Fig. 5(d)) with respect to the LSDA value of $E_{B}\sim$0.96 eV. At last, the $\Delta_{7}^{\downarrow}$ peak is shifted to $E_{B}\sim$1.20 eV in the experiment (Fig. 5(c)) and to $E_{B}\sim$1.22 eV in the LSDA+DMFT+1SM calculation (Fig. 5(d)) compared to $E_{B}\sim$1.38 eV in the LSDA calculation. This picture is consistent with an averaged experimental mass enhancement $m^{*}/m_{0}\approx$1.26, which should be compared to the theoretical value of $m^{*}/m_{0}\approx$1.29 for U=2.5 eV. A similar analysis can be carried out for the LSDA+DMFT+1SM spectra shown in Fig. 5(b) and Fig. 5(d) for U values around 2.5 eV. In particular, we obtain $m^{*}/m_{0}\approx$1.22 for U=2 eV in average, and $m^{*}/m_{0}\approx$1.55 for U=3 eV. Although the deviations between the experimental and theoretical mass enhancement factors for different U values are not large, from this evaluation U=2.5 eV is closest to the experiment. However, a precise determination of U requires a more sophisticated approach based on a full analysis of this type in the whole Brillouin zone. This was done in order to decide the best choice of U in the present calculations: the complete set of experimental and LSDA+DMFT+1SM spectra were directly compared in terms of BE position, intensity and width of the peaks over the whole Brillouin zone for different values of U, and better overall agreement was found for U=2.5 eV. Another criterion which was also combined to the one just mentioned is related to the spin character of the different bands. An example showing this can be noticed in e.g., Fig. 5(a) and Fig. 5(b), where the LSDA+DMFT+1SM spin-resolved spectra for U=2 eV exhibit a majority spin peak at $E_{B}\sim$0.6 eV which in turn, is at the same BE as the experimental minority spin peak of $\Delta_{8,9}^{\downarrow}$ symmetry.

Furthermore, it can be seen in Fig. 5(b) and Fig. 5(d) that increasing the value of U does not shift the peaks significantly towards the Fermi level. This can be attributed to the following reasons: (i) Most of the intensity in the spectra for BE$\leq$0.5 eV is due to surface-related features, which are not strongly sensitive to changes in the U parameter, (ii) the shift of pure bulk-like states only occurs in a narrow energy interval from 0.5 to 1.8 eV, otherwise their sensitivity to U is reduced through coupling to surface-related states near the border of the gap, (iii) there are no strong changes in $\Sigma_{DMFT}$ in the region very close to the Fermi level. Because of these reasons, in Co the peaks exhibit smaller shifts with increasing U than in, e.g. Fe, because they are closer to the Fermi level, and most of the bulk bands are located in a narrow energy interval. Besides, the coupling between surface and bulk states also plays a role, since in Co there is a gap around the $\Gamma$ point. This is also related to the fact that due to the hexagonal structure the bands are relatively flat, leading to real surface states which are less sensitive to changes in U and a weak bulk-surface coupling.
\subsection{Spin-dependent quasiparticle lifetimes}

The last issue to be discussed is the quality of the agreement between the experimental and theoretical linewidths of the Co photoemission peaks. As it was pointed out above, it is clear that the Co calculations presented here underestimate the scattering rates, similar to what we found in Fe.\cite{Sanchez-Barriga-PRL-2009} In Fig. 5, it can also be noticed that the width of the peaks does not increase with U. In the narrow BE range we are focusing on, increasing U shifts the peaks to lower BE and also leads to nearly-constant values of the energy-dependent Im$\Sigma_{DMFT}$. Besides, since the effect of U on the peak positions is relatively small, a change in the linewidth can hardly be appreciated. This holds for both bulk-like and surface-related features. 

We emphasize that the additional experimental broadening cannot be caused by final state effects \cite{Smith-PRB-93}, since (i) their contribution to the linewidth of surface states can be neglected because no dispersion normal to the surface exists in this case, (ii) they can be neglected when compared to the experimental resolution near the critical points (e.g. at the $\Gamma$ and A points in Fig. 5), (iii) they are completely considered in the 1SM calculations, (iv) they are expected to be small for Co(0001) because due to the hexagonal structure the bands do not show strong initial state dispersion. Because of these reasons, we exclude final state effects as possible mechanism for the extra broadening observed in the experiment. 

Nevertheless, in the following we would like to obtain more quantitative information on the discrepancy between experiment and theory we have observed concerning the linewidths. As mentioned before, a fitting procedure was used to extract the experimental peak positions from SARPES experiments shown in Fig. 3(a) and Fig. 3(b). This evaluation is also useful for an estimation by how much the theoretical linewidths are underestimated compared to the experimental results. The use of spin-resolution is remarkably important in this case because from the spin-dependent behaviour of the self-energy in the theoretical calculations one would expect the experimental linewidths of initial states to be spin-dependent as well. 

 The experimental spin-resolved spectra were fitted at various k values by a sum of Lorentzians plus a background. The fitting procedure works as follows: for a particular k-point, a SARPES spectrum containing N peaks is fitted by a function involving a convolution of the form:
\begin{equation}
\begin{split}
&I_{k}^{\uparrow,\,\downarrow}(E)=\\
&=\left(f(E,T)\cdot\sum_{i=1}^{N}M^{2}_{i}\cdot A_{k\,i}(E_{i},\omega_{i})+B_{k}(E)\right)\otimes G(h\nu)
\end{split}
\label{eq:conv}
\end{equation}
where $f(E,T)$ is the Fermi function and $E_{i}$, $\omega_{i}$ and the matrix-elements $M_{i}$ are fitting parameters corresponding to the BE, width and intensity of the quasiparticle peaks for different polarizations, respectively. The spectral function $A_{k\,i}(E_{i},\omega_{i})$ is approximated by Lorentzian functions and $B_{k}(E)$ assumed to be a Shirley-like background \cite{Shirley-PRB-1972}. Here the subindex $k$ denotes the ${\bf k}_{\perp}$ points. The full width at half maximum (FWHM) of the Gaussian slit function $G(h\nu)$ corresponds to the total energy resolution $\Delta E_{tot}$ of the experiment, which is photon energy-dependent, i.e.:
\begin{equation}
\Delta E_{tot}=\sqrt{(\Delta E_{a})^2+(\Delta E_{h\nu})^2}
\end{equation}
where $\Delta E_{a}$ and $\Delta E_{h\nu}$ are the energy resolutions of the electron energy analyzer and beamline monochromator, respectively. The Shirley-like background $B_{k}(E)$, which accounts for the effect of the inelastic scattering of electrons, is calculated as:
\begin{equation}
B_{k}(E)=C\cdot\left(\int_{E_{0}}^{\infty}dE\cdot \sum_{i=1}^{N}M^{2}_{i}\cdot A_{k\,i}(E_{i},\omega_{i})\right)
\label{eq:background}
\end{equation}
where $C$ is a fitting constant and the integration is done over the complete spectrum with $E$ as a running parameter over the kinetic energy axis. 

\begin{figure}[t!]
\centering
\includegraphics [width=0.4\textwidth]{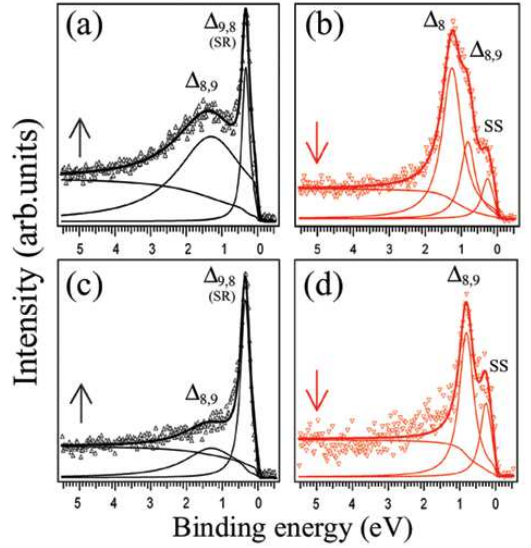}
\caption{(color online). Fits to spin-resolved valence band spectra at x=0.62$\Gamma$A (h$\nu$=56 eV) of the hcp Co(0001) bulk Brillouin zone. (a), (b) for p- and (c), (d) for s-polarization. Black (dark) [red (light)] lines are the results of the fits to majority [minority] spin spectra, which are indicated by upwards [downwards] triangles and arrows, respectively. Thin lines correspond to the different Lorentzian functions and backgrounds used in the fits. Thick lines indicate the final fit of each spectrum obtained by using Eq.~(\ref{eq:conv}). Symmetry labels, surface resonance (SR) and surface state (SS) features are indicated.}
\label{fig:Fig6}
\end{figure}

Figure 6 shows few selected results of the fits obtained by this procedure for p (Fig. 6(a) and Fig. 6(b)) and s (Fig. 6(c) and Fig. 6(d)) polarized photons, and for both majority (Fig. 6(a) and Fig. 6(c)) and minority (Fig. 6(b) and Fig. 6(d)) spin states. We should keep in mind that such an evaluation is problematic due to the strong energy-dependent behaviour of Im$\Sigma$ which leads to the formation of asymmetric Lorentzians. The appearance of such an asymmetry can be associated with the damping of quasiparticle excitations and the corresponding increase of the incoherent part of spectral function with increasing BE. However, at energies close enough to the Fermi level, we would expect this asymmetry to be strongly reduced since most of the peaks are due to coherent excitations which correspond to quasiparticles with well-defined energy and momentum.

For the fitting procedure we have adopted an experimental approach: almost degenerated bands which cannot be resolved due to the experimental energy resolution are fitted with a single peak. An exception to this point occurs when some bands can be suppressed due to matrix elements effects. In this context, polarization-dependent measurements are suited to accurately obtain the linewidth of specific bands that are nearly degenerated and cannot be resolved independently under certain light polarization conditions. This is the case of e.g., the majority spin states of $\Delta_{8,9}^{\downarrow}$ symmetry located at $E_{B}\sim$1.4 eV for p-polarization data (Fig. 6(a)), which form a shoulder in the corresponding s-polarization spectra around the same BE (Fig. 6(b)). The same holds for the $\Delta_{8,9}^{\downarrow}$ minority spin states located at $E_{B}\sim$0.7 eV for s-polarization data (Fig. 6(d)), which in turn form a shoulder in the corresponding p-polarization spectra (Fig. 6(b)). This allows us to fix the width and BE positions obtained from the fits of individual bands for a given polarization in order to fit the shoulders of almost degenerated bands which are observed with the opposite polarization.

Concerning the linewidths, the result of this analysis is presented in Fig. 7, where we have extracted the spin-dependent experimental Im$\Sigma_{exp}$ of Co as a function of BE and k and compared it to the theoretical one. The results for majority (minority) spin electrons are shown in Fig. 7(a) (Fig. 7(b)), together with the corresponding LSDA+DMFT (U=2.5 and 3 eV) and LSDA+3BS calculations (U=2.5 eV). It is clear that no quantitative agreement between the theoretical calculations and the experimental data is observed. 
The experimental data points have been corrected by all the non-electronic contributions to the broadening with the exception of impurity scattering (Im$\Sigma_{imp}=\Gamma_{imp}$/2, where $\Gamma$=FWHM) since it can only be determined once all the other contributions have been corrected for. 

\begin{figure}[t!]
\centering
\includegraphics [width=0.5\textwidth]{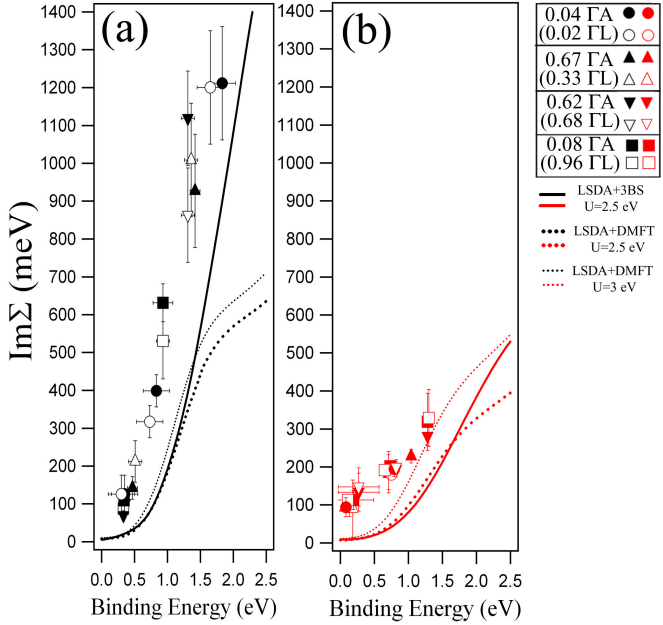}
\caption{(color online). Comparison between the experimental (symbols) and theoretical (lines) imaginary part of the self-energy of hcp Co(0001) for (a) majority (black (dark) color) and (b) minority (red (light) color) spin electrons. The experimental data points only contain the electronic and impurity scattering contributions to the linewidth. On the right hand side, labels to identify the different k-points are also given for both hcp and fcc lattices. Full and open symbols are for p- and s-polarization, respectively. The theoretical calculations correspond to Im$\Sigma_{DMFT}$ for U=2.5 eV (thick dotted lines) and U=3 eV (thin dotted lines) and to Im$\Sigma_{3BS}$ for U=2.5 eV (solid lines).}
\label{fig:Fig7}
\end{figure}
The theoretical calculations, on the other hand, only contain the electronic contribution. Since the broadening due to energy resolution is already included in the fitting procedure, the other experimental corrections are due to electron-phonon broadening ($\Gamma_{e-ph}$) and final-state broadening ($\Gamma_{f}$), where $\Gamma$=2Im$\Sigma$ is the scattering rate. In general, all these contributions add linearly and give a total scattering rate $\Gamma_{t}$ such that
\begin{equation}
\Gamma_{t}=\Gamma_{e-e}+\Gamma_{e-ph}+\Gamma_{imp}+\Gamma_{f}
\end{equation}
where $\Gamma_{e-e}$ indicates the contribution to the linewidth due to electron-electron scattering. In the Debye model and for the high temperature limit \cite{Aschcroft, Fedorov-PRB-2002}, the electron-phonon broadening depends linearly on the temperature $T$ as: 
\begin{equation}
\Gamma_{e-ph}=2Im\Sigma_{e-ph}=2\pi\lambda k_{B}T
\label{eq:e-ph}
\end{equation}
where $k_{B}$ and $\lambda$ are the Boltzmann and electron-phonon coupling constants, respectively. Eq. (\ref{eq:e-ph}) holds for temperatures $T\geq T_{Debye}/3$ (here T=300 K and $T_{Debye}$(Co)=385 K). From resistivity measurements\cite{Allen-PRB-1987} we obtain for Co $\lambda\approx 0.3$ and as a result Im$\Sigma_{e-ph}\approx$ 25 meV.

The final state broadening was determined by unfolding the bands around the A point and calculating the experimental electron initial and final state velocities assuming a free electron parabola as final state. Firstly, the initial state velocities were calculated from the derivative of the E(${\bf k}_{\perp}$) experimental dispersions of the different majority and minority spin bands. Secondly, the inelastic mean free path values were taken from the semiempirical estimation given by Penn \cite{Penn-PRB-1987} and available reference data.\cite{Werner-JPCRD-2009} In general, due to the relatively flat dispersions of the d bands, a maximum ratio between the initial and final state group velocities $|v_{i\perp}/v_{f\perp}|\approx 0.04$ was obtained, equivalent to a maximum contribution to Im$\Sigma_{exp}$ of about 40 meV. This indicates that final-state effects are not the main broadening mechanism, as mentioned above. Finally, the electron-impurity scattering, an energy and temperature-independent quantity, can be directly estimated as an average for both spin channels. It corresponds to the value of Im$\Sigma_{exp}$ exactly at the Fermi level position, and by extrapolation it amounts to Im$\Sigma_{imp}\sim$50 meV.

Although no quantitative agreement between the experimental and theoretical Im$\Sigma$ is obtained for Co, both spin channels are qualitatively reproduced in the theoretical results. Interestingly, the experimental data support the existence of a pronounced spin-dependent effect in the quasiparticle lifetimes in agreement with previous observations.\cite{Monastra-PRL-2002} However, this is the first time that the use of spin resolution has been exploited to demonstrate such a notorious effect. In the experimental results, we obtain to a first approximation a k-independent Im$\Sigma_{Exp}$ which exhibits an almost linear energy-dependent behaviour approximated as Im$\Sigma_{Exp}^{\uparrow}\sim 0.84E$ and Im$\Sigma_{Exp}^{\downarrow}\sim 0.16E$ for majority and minority spin electrons, respectively. Furthermore, the relative overall agreement between the experiment and the theoretical results indicates that in the case of Co, the role of non-local correlations is not as important as for Fe.\cite{Sanchez-Barriga-PRL-2009} This could be attributed to a reduction of the non-locality of correlation effects with increasing atomic number.

Regarding the theoretical calculations of Im$\Sigma_{DMFT}$ in Fig. 7, it can be seen that increasing U does not strongly affect the global behaviour of the calculated curves. Note that this is something expected following our previous discussion concerning the shifting of the peaks with increasing U, which was also not very pronounced. In this respect, it should be emphasized that our results do not imply a violation of the Kramers-Kronig consistency \cite{SH,Hufner-highres} which holds between the real and imaginary parts of the self-energy, neither in the experiment nor in the theoretical calculations. This is because the Kramers-Kronig consistency results from a global transformation over an extremely large BE range, which runs up to infinity. This means that in the narrow BE energy interval studied here, the deviations between the experimental and theoretical self-energies may become substantial without any noticeable effect in the global behaviour of the Kramers-Kronig transformation. To check this more in detail experimentally, time-consuming measurements over an extremely wide BE range would be necessary. In Fig. 7, the deviations between experimental and theoretical data seem to be equally pronounced for both majority and minority spin electrons, with the exception of the minority spin Im$\Sigma_{DMFT}$ for U=3 eV, which seems to agree better if the electron-impurity scattering contribution is subtracted from the experimental curve. Nevertheless the overall agreement shows that the experimental Im$\Sigma$ is roughly a factor of 2 too large in average when compared to the calculations. In the LSDA+DMFT calculations, this discrepancy is more pronounced in the majority spin channel at higher binding energies than at lower ones. Note that both the Im$\Sigma_{DMFT}$ and Im$\Sigma_{3BS}$ agree well for the same U values, with the exception that at higher binding energies they start to deviate from each other. We attribute this difference to the fact that LSDA+3BS calculations explicitly include electron-hole pair excitations which increase Im$\Sigma$. The extent of the present agreement contrasts a very recent non spin-resolved analysis from Moulazzi et al. \cite{Mulazzi-PRB-2009}, in which excellent agreement between experiment and previously reported results of $Im\Sigma_{DMFT}$ \cite{Grechnev-PRB-2007} is achieved for both spin channels, most probably due to an arbitrary linewidth assignment of nearly-overlapping spin contributions.

\section{Conclusion}

In summary, we have shown a detailed comparison between spin and angle-resolved photoemission experiments and state-of-the-art theoretical calculations of ferromagnetic hcp Co(0001). We have analyzed in detail the agreement between theoretical many-body 3BS and DMFT based calculations and experimental spin-resolved photoemission spectra. The theoretical methods lead to an important improvement compared to LSDA calculations by including, in an unified picture, many-body corrections with multiple-scattering, matrix-elements and surface-related effects. However, we have demonstrated that these theories, at least in the current implementation, do not find exact quantitative agreement with the experiment, in particular concerning the scattering rates. Although further theoretical work is needed to determine the origin of these differences, a possible explanation could be related to the existence of non-local correlation effects. Such non-locality is caused by long-range electron-electron interactions, which are excluded from the Hubbard model. These non-local correlation effects would contribute with an additional k-dependent correction to the self-energy functions $\Sigma_{DMFT}$ or $\Sigma_{3BS}$. This would result in second-order variations in the calculated Re$\Sigma$ and Im$\Sigma$, which may lead to a much better quantitative agreement between theory and experiment. The fact that non-local correlation effects can have a certain influence on the intermediate energy states is closely related to what we found for Fe.\cite{Sanchez-Barriga-PRL-2009} However, the better overall agreement between theory and experiment obtained for Co at intermediate U values indicates that non-local correlations become weaker with increasing atomic number. It would be very interesting to apply similar investigations to itinerant magnets like the series FeSi, MnSi and CoSi, where we also expect the non-local character of correlations effects to play an important role. Although non-local interactions are beyond the many-body theories used in this work, a promising method to overcome this limitation in the near future is the {\it dual fermion} approach.\cite{Rubtsov-PRB-2008} Alternatively, it would be an important step to implement into realistic electronic structure calculations the recently proposed parameter-free extended DMFT+GW scheme \cite{Biermann-PRL-2003}, in which both the on-site and off-site correlations are included. 
\section{Acknowledgements}
Financial support by the Deutsche Forschungsgemeinschaft through FOR 1346 and MI-1327/1 is gratefully acknowledged.

$^{*}$Electronic address: sbarriga@bessy.de
\end{document}